\documentclass[12pt]{iopart}
\usepackage{iopams}
\usepackage{graphicx}
\usepackage{color}

\begin{document}

\title[Quasiclassical Asymptotics and Coherent States for Bounded Discrete Spectra]
{Quasiclassical Asymptotics and Coherent States for Bounded Discrete Spectra}
\author{K. G\'{o}rska$^{a, b}$, K. A. Penson$^{b}$, A. Horzela$^{c}$, G. H. E. Duchamp$^{d}$, P. Blasiak$^{c}$ and A. I. Solomon$^{b, e}$ \vspace{2mm}}

\address
{$^a$ Nicolaus Copernicus University, Institute of Physics, ul. Grudzi\c{a}dzka 5/7,\\
PL 87-100 Toru\'{n}, Poland\vspace{2mm}}

\address
{$^b$ Laboratoire de Physique Th\'eorique de la Mati\`{e}re Condens\'{e}e,\\
Universit\'e Pierre et Marie Curie, CNRS UMR 7600\\
Tour 13 - 5i\`{e}me \'et., B.C. 121, 4 pl. Jussieu, F 75252 Paris Cedex 05, France\vspace{2mm}}

\address
{$^c$ H. Niewodnicza\'nski Institute of
Nuclear Physics, Polish Academy of Sciences\\
ul. Eliasza-Radzikowskiego 152,  PL 31342 Krak\'ow, Poland\vspace{2mm}}

\address
{$^d$ Universit\'e Paris XIII, LIPN, Institut Galil\'{e}e, CNRS UMR 7030, \\
99 Av. J.-B. Clement, F 93430 Villetaneuse, France \vspace{2mm}}

\address
{$^e$ The Open University, Physics and Astronomy Department\\
Milton Keynes MK7 6AA, United Kingdom\vspace{2mm}}

\eads{\linebreak \mailto{kasia\_gorska@o2.pl},
\mailto{penson@lptl.jussieu.fr},
\mailto{andrzej.horzela@ifj.edu.pl},
\mailto{ghed@lipn-univ.paris13.fr},
\mailto{pawel.blasiak@ifj.edu.pl},
\mailto{a.i.solomon@open.ac.uk}}


\begin{abstract}
\\
We consider discrete spectra of bound states for  non-relativistic motion in attractive potentials $V_{\sigma}(x) = -|V_{0}|\, |x|^{-\sigma}$, $0<\sigma\leq 2$. For these potentials the quasiclassical approximation for $n\to \infty$ predicts quantized energy levels $e_{\sigma}(n)$ of a bounded spectrum varying  as $e_{\sigma}(n) \sim -n^{-2\sigma/(2-\sigma)}$. We construct collective quantum states using the set of wavefunctions of the discrete spectrum assuming this asymptotic behaviour. We give examples of states that are normalizable and satisfy the resolution of unity, using explicit positive functions. These are coherent states in the sense of Klauder and their completeness is achieved via exact solutions of  Hausdorff moment problems, obtained by combining Laplace and Mellin transform methods. For $\sigma$ in the range $0<\sigma\leq 2/3$ we present exact implementations of such states for the parametrization $\sigma = 2(k-l)/(3k-l)$, with $k$ and $l$ positive integers satisfying $k>l$.
\end{abstract}

\maketitle

\section{Introduction}
The construction of collective quantum states characterizing  the whole spectrum of a quantum system is a challenging problem which  depends sensitively on the nature of the potential involved. The standard coherent states (CS) are closely related to the harmonic oscillator potential and are defined, for complex $z$, by
\begin{equation}\label{eqI1}
|z\rangle = \exp(-|z|^2/2)\, \sum_{n=0}^{\infty} \frac{z^n}{\sqrt{n!}} \, |n\rangle,
\end{equation}
where $\{|n\rangle\}_{n=0}^{\infty}$ is the Fock space of eigenfunctions of operator $H = p^2/(2 m) + m\omega^2\,x^2/2$ satisfying $H|n\rangle = \hbar\omega(n+1/2)|n\rangle$, $\langle n|n'\rangle = \delta_{n, n'}$. There have been  many attempts to generalize the construction of Eq.~(\ref{eqI1}) for potentials other than $\sim x^2$ \cite{MMNieto80, JRRay82}. These efforts were  hampered by the fact that the exact eigenstates and spectra of general potentials $V(x)$ are known in only a few cases, and for a very special form of $V(x)$. In fact for purely power-law type potentials of the form $V_{\sigma}(x) = \pm |V_{0}|\, |x|^{-\sigma}$ the only exact solutions we know are for $|\sigma| = 1, 2$. The attractive case $\sigma > 2$ is considered in a certain sense as unphysical \cite{LDLandau00}. This paucity forces one either to resort to approximations or to construct trial generalizations of (\ref{eqI1}) in which some known ingredients are built in, whereas other features are not accounted for. In this context, the quasiclassical approach may play a prominent role \cite{PMoxhay80, CQuigg79, VGalitski85, KBerrada10}.

In this work we will be concerned with attractive power-law potentials and the discrete part of their spectrum, which is known to be bounded \cite{LDLandau00} and which we assume here to be nondegenerate. Here $V_{\sigma}(x) = -|V_{0}|\,|x|^{-\sigma}$, $0<\sigma\leq 2$ and the quasiclassical estimate for the spectrum $E_{\sigma}(n)$ is obtained from the Bohr-Sommerfeld quantization rule
\begin{equation}\label{eqI2}
\frac{1}{\hbar} \int_{a}^{b} \sqrt{2m\left[E_{\sigma}(n) - V_{\sigma}(x)\right]}\,dx = \pi (n + 1/2), \quad n =0, 1, \ldots
\end{equation}
which can be evaluated as (with $\hbar = 1$)
\begin{equation}\label{eqI3}
E_{\sigma}(n) = - \left(\frac{\pi}{2}\, \frac{n+1/2}{\sqrt{2m}\, D(\sigma)}\, |V_{{0}}| \right)^{-2\sigma/(2-\sigma)} \, \to \, - n^{-2\sigma/(2-\sigma)}, \,\, n\to \infty,
\end{equation}
see \cite{VGalitski85} for derivation and \cite{PMoxhay80, CQuigg79} for various refinements. In Eq.~(\ref{eqI2}) $x=a$ and $x=b$ are turning points of the potential, defined by $p(a) = p(b) = 0$, and $D(\sigma) = \int_{a}^{b}\sqrt{x^{\sigma} - 1} dx$.

We now use dimensionless units in which $H_{\sigma}(p, x) \Rightarrow h_{\sigma}(p, x)~=~p^2 - x^{-\sigma}$. Then, we fix the value $e_{\sigma}(0) = 0$ and thus arrive at the quasiclassical form of the spectrum
\begin{equation}\label{eqI4}
e_{\sigma}(n) \approx 1 - c\, n^{-2\sigma/(2-\sigma)} + \ldots, \quad 0<\sigma\leq 2, \, n\to\infty,
\end{equation}
where the constant $c>0$. We shall now follow the approach developed in \cite{JPGazeau99, JRKlauder01, DPopov08} and incorporate the form of Eq.~(\ref{eqI4}) to construct the generalization of states defined in Eq.~(\ref{eqI1}) which are specially adapted to attractive potentials proportional to $- |x|^{-\sigma}$. (For a recent treatment of CS for continuous spectra, see \cite{JPGazeau99, JBenGeloun09}.)

To do so we use the method elaborated for discrete spectra and propose the specific form of a collective quantum state spanned by a set of eigenfunctions $|n, \sigma\rangle$ of $h_{\sigma}(p, x)$ satisfying (with $\langle n, \sigma| n', \sigma\rangle = \delta_{n, n'}$) asymptotically, as $n\to\infty$
\begin{equation}\label{eqI5}
h_{\sigma}(p, x)|n, \sigma\rangle = e_{\sigma}(n)|n, \sigma\rangle.
\end{equation}
Note that the equality sign in Eq.~(\ref{eqI5}) is in general not valid for small values of $n$. Although $|n, \sigma\rangle$ are not known in general, we still construct a trial wave function \cite{JPGazeau99, JRKlauder01, AHElKinani02, JPAntoine01} in the form generalizing Eq.~(\ref{eqI1}):
\begin{equation}\label{eqI6}
|J, \gamma, \sigma\rangle = \mathcal{N}_{\sigma}^{-1/2}(J)\, \sum_{n=0}^{\infty} \frac{J^{n/2}\exp(-i\,\gamma\,e_{\sigma}(n))}{\sqrt{\rho_{\sigma}(n)}}\,|n, \sigma\rangle,
\end{equation}
where $J>0$, and $\gamma$ are real and $\rho_{\sigma}(n)$ are so chosen as to assure the convergence of the normalization $\mathcal{N}_{\sigma}(J)>0$,
\begin{equation}\label{eqI7}
\mathcal{N}_{\sigma}(J) = \sum_{n=0}^{\infty} \frac{J^n}{\rho_{\sigma}(n)}\,<\,\infty,  \quad 0\leq J\leq R\leq \infty.
\end{equation}
The choice of $\rho_{\sigma}(n)$ in Eq.~(\ref{eqI6}) is dictated by an "action identity" of the form \cite{JPGazeau99, JRKlauder01}
\begin{equation}\label{eqI8}
\langle J, \gamma, \sigma| h_{\sigma}(p, x) | J, \gamma, \sigma\rangle \,=\, J
\end{equation}
which directly implies
\begin{equation}\label{eqI9}
\rho_{\sigma}(n) = \prod_{j=1}^{n} e_{\sigma}(j), \quad \rho_{\sigma}(0) = 1,
\end{equation}
and  consequently the basic relation follows:
\begin{eqnarray}\label{eqI10}
e_{\sigma}(n) &=& \frac{\rho_{\sigma}(n)}{\rho_{\sigma}(n-1)}, \quad n=1, 2, \ldots, \\[0.7\baselineskip] \nonumber
e_{\sigma}(0)&=&0,
\end{eqnarray}
which, within our approach should be understood in the asymptotic sense. The states $|J, \gamma, \sigma\rangle$ should satisfy the resolution of identity with a weight function $W_{\sigma}(J)>0$:
\begin{equation}\label{eqI11}
\int\,dJ\, d\gamma\,\, |J, \gamma, \sigma\rangle W_{\sigma}(J)\langle J, \gamma, \sigma| \, =\, \sum_{n=0}^{\infty} |n, \sigma\rangle\langle n, \sigma| = \mathbb{I}
\end{equation}
which reduces ({\it vide} Eq. (102) of Ref. \cite{JRKlauder01}) to an infinite set of integral equations for an unknown positive function $W_{\sigma}(x)$:
\begin{equation}\label{eqI12}
\int_{0}^{R} x^n \,\left[\frac{W_{\sigma}(x)}{\mathcal{N}_{\sigma}(x)} \right]\, dx = \int_{0}^{R}\, x^n\, \widetilde{W}_{\sigma}(x)\, dx = \rho_{\sigma}(n), \,n=0, 1, \ldots
\end{equation}
where $\rho_{\sigma}(0) = 1$. If $R<\infty$, Eqs.~(\ref{eqI12}) is the Hausdorff moment problem \cite{NIAkhiezer65}. It is known that if  for a given set of $\rho(n)$'s the positive solution of Eqs.~(\ref{eqI12}) exists then it is always unique \cite{NIAkhiezer65}. The situation is very different for Hamiltonians with unbounded discrete spectra which lead to the Stieltjes moment problem with $R=\infty$ in Eq.~(\ref{eqI12}). In this case the solutions can be either unique or non-unique, see \cite{KAPenson10}. Observe that apart from their orthogonality no specific knowledge of the $|n, \sigma\rangle$'s is required to derive Eq.~(\ref{eqI12}).

The state $|J, \gamma, \sigma\rangle$ defined by Eq.~(\ref{eqI6}), with $\rho_{\sigma}(n)$ satisfying Eqs.~(\ref{eqI10}) and (\ref{eqI12}), is a generalized CS state in a sense of Klauder \cite{JRKlauder01}, asymptotically relevant for $V_{\sigma}(x)\sim~-|x|^{-\sigma}$.


\section{Generating solutions of Hausdorff moment problems}

Our strategy from now on is: a) identify the form of $\rho_{\sigma}(n)$'s to assure that Eq.~(\ref{eqI12}) can be solved for $\widetilde{W}_{\sigma}(x)>0$; b) calculate the associated energy spectrum from Eq.~(\ref{eqI10}); and c) identify the exponent $\sigma$ obtained from Eq.~(\ref{eqI4}) and thus link the potential $V_{\sigma}(x)$ to $\rho_{\sigma}(n)$ and $|J, \gamma, \sigma\rangle$. Evidently the correspondence in c) above is not unique: one may give different $\rho_{\sigma}^{(r)}(n)$, $r=1, 2, \ldots$ which  yield the same asymptotics via Eq.~(\ref{eqI4}).

We consider Eq.~(\ref{eqI12}) as a Mellin transform $f^{*}(s) = \mathcal{M}[f(x); s] = \int_{0}^{\infty} x^{s-1} f(x) dx$, $s$ complex \cite{IASneddon72}:
\begin{equation}\label{eqI13}
\mathcal{M}\left[\widetilde{W}_{\sigma}(x); s\right] = \rho_{\sigma}(s-1), \quad s\geq 1,
\end{equation}
or equivalently
\begin{equation}\label{eqI14}
\widetilde{W}_{\sigma}(x) = \mathcal{M}^{-1}\left[\rho_{\sigma}(s-1); x\right],
\end{equation}
where $\mathcal{M}^{-1}$ is the inverse Mellin transform. We observe that the moments used in this work will always be of such a nature that the integration range $R$ in Eqs.~(\ref{eqI12}) will be equal to $1$. Then the moment sequences will be decreasing functions of $n$. (We stress that this  is not a general rule: there exist Hausdorff moment problems necessitating $R>1$ for which the moment sequences are increasing \cite{KAPenson, KAPenson01}.)

In our search for positive solutions of Eqs.~(\ref{eqI12}) we were greatly helped by a relation between the Laplace transform and a special case of Mellin transform \cite{FOberhettinger74}. To elucidate this link suppose that a function $F(x)$ is considered for which its Laplace transform is known:
\begin{equation}\label{eqI15}
\mathcal{L}[F(x); p] \equiv F(p) = \int_{0}^{\infty} e^{-px} \, F(x)\, dx, \,\, p>0.
\end{equation}
We now perform  a change of variable $x = \ln(1/y)$ in Eq.~(\ref{eqI15}), which gives
\begin{equation}\label{eqI16}
F(p) = \int_{0}^{\infty} y^{p-1}\, F(\ln(1/y))\, H(1-y)\, dy,
\end{equation}
where $H(z)$ is the Heaviside function. Through a formal renaming $p \leftrightarrow s$ we treat Eq.~(\ref{eqI16}) as a Mellin transform
\begin{equation}\label{eqI17}
F(s) = \mathcal{M}[F(\ln(1/x))\,H(1-x); s].
\end{equation}
The relations of Eqs.~(\ref{eqI15})-(\ref{eqI17}) allow one to search for possible solutions of the Hausdorff moment problem (\ref{eqI12}) for $R=1$ via the method of the inverse Laplace transform \cite{APPrudnikov92}, with a succession of following steps: i)~choose a strictly decreasing sequence of moments $\rho(n)$, $n=0, 1, \ldots$; ii)~rename them as $F(p+1)$, s.Eq.~(\ref{eqI16}); iii) search for the inverse Laplace transform $F(x)$ corresponding to $F(p)$ and check whether $\widetilde{W}(x) = F(\ln(1/x))\,H(1-x)$ is a positive function on $[0, 1]$; then, if so, $\widetilde{W}(x)$ is the solution of the Hausdorff moment problem Eqs.~(\ref{eqI12}). One is helped here  by the fact if that $F(x)$ is positive on $[0, \infty)$ then $F(\ln(1/x))$ is positive on $[0, 1]$.

\subsection{Illustrative example}

We illustrate this approach with an example directly related to our construction. We choose the moments as $\rho(n) = e \exp(- \sqrt{n+1})$, $\rho(0) = 1$; the relabelling $\rho(n) \leftrightarrow F(p+1)$ gives $F(p) = e \exp( - \sqrt{p})$ which, with the formula 2.2.1.9 on p.~52 of \cite{APPrudnikov92} for $a=1$, is the Laplace transform
\begin{equation}\label{eqI18}
e \exp(-\sqrt{p}) = \int_{0}^{\infty} e^{-p x} \left[\frac{e}{2\sqrt{\pi}\, x^{3/2}}\, \exp(- 1/4x)\right] \, dx.
\end{equation}
In the next step we verify that $\frac{e}{2\sqrt{\pi}} [\ln(1/x)]^{-3/2}  \exp\left(-\frac{1}{4\ln(1/x)}\right)$
is a positive function on $[0, 1]$ and consequently
\begin{equation}\label{eqI19}
\int_{0}^{1} x^n \, \left[\frac{e}{2\sqrt{\pi}\, [\ln(1/x)]^{3/2}}\,\exp\left(-\frac{1}{4\ln(1/x)}\right) \right] \, dx \, =\, e \cdot e^{-\sqrt{n+1}},
\end{equation}
$n=0, 1, \ldots$, is a complete solution of the Hausdorff moment problem Eqs.~(\ref{eqI12}). With the above moments the spectrum Eq.~(\ref{eqI10}) is
\begin{equation}\label{eqI20}
e(n) = e^{\sqrt{n} - \sqrt{n+1}},
\end{equation}
and its $n\to\infty$ asymptotics is
\begin{equation}\label{eqI21}
e(n) \to 1 - \frac{1}{2\,n^{1/2}} + \frac{1}{8\, n} + \ldots,
\end{equation}
which with Eq.~(\ref{eqI4}) determines $\sigma = 2/5$ and $c=1/2$. The generalized coherent state describing such a spectrum is given by Eq.~(\ref{eqI6}) with the normalization
\begin{equation}\label{eqI22}
\mathcal{N}(J) = \frac{1}{e}\,\sum_{n=0}^{\infty} e^{(n+1)^{1/2}}\, J^n, \quad 0\leq J < 1.
\end{equation}
The CS $|J, \gamma, 2/5\rangle$ is then asymptotically relevant for motion in the potential $V_{2/5}(x)\sim - |x|^{-2/5}$. The formula Eq.~(\ref{eqI19}) can also be cross-checked by referring to the tables of inverse Mellin transforms (see formula 3.7 on p. 174 for $\alpha = 1$ of Ref. \cite{FOberhettinger74}). In the spirit of  Eqs.~(\ref{eqI12}) we call the weight function in Eq.~(\ref{eqI19}) $W_{2/5}(x)$, $0\leq x\leq 1$, which can also be derived from $e L(1/2, \ln(1/x))$, where $L(\gamma, x)~=~\sqrt{\frac{\gamma}{2\pi}}\, x^{-3/2}\, e^{-\gamma/(2x)}$ is the so-called one-sided L\'{e}vy stable distribution \cite{JPKahane94}.

We now present  a more general case which can be treated with the above method. For that purpose we again stress  that we are looking for special sequences of moments $\rho(n)$ satisfying the Hausdorff moment equations which  at the same time possess very specific asymptotic properties implied by Eqs.~(\ref{eqI4}) and (\ref{eqI10}). These are very restrictive conditions indeed. The search for such solutions may proceed by exclusion and at the beginning it was not certain at all if a general solution existed. It is all the more satisfying that a parametrization can be given that produces a vast ensemble of solutions, at least for some range of values of $\sigma$.

\subsection{Full solutions for $\sigma$ rational in the range $0 <\sigma\leq 2/3$ and for $\sigma = 1$}

Let us define a sequence of moments, parametrized by $a$, $\nu, k$ and $l$, and given by:
\begin{equation}\label{eqI23}
\rho^{(a, \nu)}(k, l; n) \,=\, e^a\,(n+1)^{-\nu}\, e^{-a\,(n+1)^{l/k}}, \,\, n=0, 1, \ldots,
\end{equation}
with conditions: $k$ and $l$ positive integers; $k>l$; $a>0$ and $\nu\geq 0$. The last condition assures the positivity of the weight function, see Appendix A. We use now the formula 2.2.1.19 listed without proof on p. 53 of \cite{APPrudnikov92}:
\begin{equation}\label{eqI24}
\int_{0}^{\infty} \, e^{-px}\, \left[\frac{\sqrt{k}\,\,l^{1/2 - \nu}}{(2\pi)^{(k-l)/2}}\, x^{\nu -1}\, G^{k, 0}_{l, k}\left(\frac{a^k\, l^l }{k^k \, x^l} \, \Big\vert\,^{\Large{\Delta(l, \nu)}}_{\Large{\Delta(k, 0)}}\right)\right]\, dx \,=\, p^{-\nu}e^{-a\,p^{\,l/k}},
\end{equation}
for $p>0$, where $G^{m, n}_{p, q}\left(z| \cdots \right)$ is Meijer's G function \cite{OIMarichev83}. The detailed demonstration of Eq.~(\ref{eqI24}) will be given elsewhere.

We transform the Eq.~(\ref{eqI24}) with $x = \ln(1/y)$ and arrive at the expression of the type of Eq.~(\ref{eqI16}), namely
\begin{eqnarray}\label{eqI25}
&\int_{0}^{\infty}& y^{p-1} \, \left[\frac{e^{a}\, \sqrt{k}}{(2 \pi)^{(k-l)/2}}\, l^{1/2 - \nu}\, [\ln(1/y)]^{\nu-1}\right. \\[0.7\baselineskip]\nonumber
&\times& \left. G^{k, 0}_{l, k} \left(\frac{a^k \, l^l}{k^k \, [\ln(1/y)]^l} \Big\vert\,^{\Delta(l, \nu)}_{\Delta(k, 0)}\right)\, H(1-y)\right]\, dy \,=\, e^a \, p^{-\nu} \, e^{-a\,p^{\,l/k}}, \,\,\, p>0.
\end{eqnarray}
In Eqs.~(\ref{eqI24}) and (\ref{eqI25}) we use a compact notation for special lists of $k$ elements \cite{APPrudnikov92}: $\Delta(k, a)~=~\frac{a}{k}, \frac{a+1}{k}, \ldots, \frac{a+k-1}{k}$. The Meijer G function is defined as an inverse Mellin transform \cite{OIMarichev83, APPrudniko98v}:
\begin{eqnarray}\label{eqI26}
&G^{m, n}_{p, q}& \left( z \Big\vert\,^{\alpha_{1} \ldots \alpha_{p}}_{\beta_{1} \ldots \beta_{q}}\right) = \mathcal{M}^{-1} \left[\frac{\prod_{j=1}^{m}\Gamma(\beta_{j}+s)\, \prod_{j=1}^{n}\Gamma(1-\alpha_{j}-s)}{\prod_{j=m+1}^{q}\Gamma(1-\beta_{j}-s)\, \prod_{j=n+1}^{p}\Gamma(\alpha_{j}+s)} ; z\right] \\[0.7\baselineskip]\label{eqI26a}
&=& G([[\alpha_{1},\ldots, \alpha_{n}], [\alpha_{n+1},\ldots, \alpha_{p}]], [[\beta_{1}, \ldots, \beta_{m}], [\beta_{m+1}, \ldots, \beta_{q}]], z),
\end{eqnarray}
where in Eq.~(\ref{eqI26}) empty products are taken to be equal to one. In Eqs.~(\ref{eqI26}) and (\ref{eqI26a}) the parameters are subject of conditions:
\begin{eqnarray}\label{eqI27}
&& z \neq 0,\,\,  0\leq m \leq q,\,\, 0 \leq n \leq p; \\[0.7\baselineskip]\nonumber &&\alpha_{j}\in\mathbb{C}, \,\, j=1, \ldots, p;\,\,\, \beta_{j}\in\mathbb{C}, \,\, j=1, \ldots, q.
\end{eqnarray}
For a full description of integration contours in Eq.~(\ref{eqI26}), general properties and special cases of the $G$ functions see \cite{OIMarichev83, APPrudniko98v}. In Eq.~(\ref{eqI26a}) we present a transparent notation, which we will use henceforth, inspired by computer algebra \cite{Maple}. With this notation  Eq.~(\ref{eqI25}) now becomes
\begin{eqnarray}\label{eqI28}
\int_{0}^{1} &y^{p-1}& \,\left[ \frac{e^a\, \sqrt{k}\, l^{1/2-\nu}}{(2\pi)^{(k-l)/2}}\,\,  [\ln(1/y)]^{\nu-1}\right. \\[0.7\baselineskip]\nonumber
&\times& \left. G\left([[\,\,\,], [\Delta(l, \nu)]], [[\Delta(k, 0)], [\,\,\,]], \frac{a^k \, l^l}{k^k \, [\ln(1/y)]^l}\right)\right] \, dy \\[0.7\baselineskip]\nonumber
&=& e^a \, p^{-\nu} \, e^{-a p^{\,l/k}}, \,\,\, p>0,
\end{eqnarray}
or, when rewritten as the moment problem with Eq.~(\ref{eqI23}):
\begin{eqnarray}\nonumber
\int_{0}^{1} &y^n& \left[\frac{e^{a}\, \sqrt{k}\, l^{1/2-\nu}}{(2\pi)^{(k-l)/2}} \, [\ln(1/y)]^{\nu-1} \, G\left([[\,\,\,], [\Delta(l, \nu)]], [[\Delta(k, 0)], [\,\,\,]], \frac{a^k \, l^l}{k^k \, [\ln(1/y)]^l}\right)\right]\, dy \\[0.7\baselineskip]\label{eqI29}
&=& \rho^{(a, \nu)}(k, l; n) \\[0.7\baselineskip]
&=& \int_{0}^{1} y^n \, \widetilde{W}^{(a, \nu)}(k, l; y) dy,  \quad n=0, 1, \ldots, \infty. \label{eqI30}
\end{eqnarray}

Observe that in Eq.~(\ref{eqI29})  only the second and third lists of parameters are non empty in the $G$ functions ({\it vide}~the notation of Eq.~(\ref{eqI26a})), as may be inferred from the conditions of Eq.~(\ref{eqI27}).  Eqs.~(\ref{eqI29}) and (\ref{eqI30}) have a very rich ensemble of solutions which will be studied for various values of the parameters $a$, $\nu$, $k$ and $l$. The role played by $a$ and $\nu$ is fundamentally different from that played by $k$ and $l$. The asymptotic behaviour as  $n\to\infty$ of $\rho^{(a, \nu)}(k, l; n)$ does not depend on $a$ and $\nu$ and the exponent of the power-law $n$ dependence of spectra is a function of $l/k$ only:
\begin{equation}\label{eqI31}
e^{(a, \nu)}(k, l; n) = 1 - \frac{l}{k}\frac{a}{n^{(k-l)/k}} + \ldots,
\end{equation}
which, by Eqs.~(\ref{eqI3}) and (\ref{eqI4}) immediately implies that $\sigma$ may be "fine-tuned" with the parametrization
\begin{equation}\label{eqI32}
\sigma = \frac{2(k-l)}{3k - l}, \quad k>l; \,\, k, l = 1, 2, \ldots\, .
\end{equation}
Eq.~(\ref{eqI32}) confines $\sigma$ to a possible range of $0<\sigma<2/3$. In Eq.~(\ref{eqI31}) the constant $c$ of Eq.~(\ref{eqI4}) is equal to $c=al/k$. The above analysis indicates that the weights $\widetilde{W}^{(a, \nu)}(k, l; y)$ for different $a$ and $\nu$ give the same asymptotics.

We shall give examples of such an asymptotic "degeneracy" with explicit forms of $\widetilde{W}^{(a, \nu)}(k, l; y)\neq\widetilde{W}^{(a, \nu ')}(k, l; y)$ for a few $\nu\neq\nu '$. For simplicity, from now on a fixed value $a=1$ will be used for all examples derived from Eq.~(\ref{eqI24}). We shall observe the onset of complexity with increasing values of $k$ and $l$: starting with $k = 4$ and $l=1$ we leave the realm of standard special functions as then the corresponding Meijer's $G$ functions can be only converted to finite sums of generalized hypergeometric functions of type $_{p}F_{q}$. They are however available through computer algebra systems \cite{Maple} and their properties are, to a large extend, readily accessible. Since $a=1$ we shall denote $\widetilde{W}^{(1, \nu)}(k, l; y) \equiv \widetilde{W}^{(\nu)}(k, l; y)$.

\subsection{Special cases:}

\noindent
1. $k=2$, $l=1$ and $\nu=0$:
\begin{eqnarray}\nonumber
\widetilde{W}^{(0)}(2, 1; y) &=& \frac{e/\sqrt{\pi}}{\ln(1/y)} \,\, G\left([[\,\,\,], [0]], [[0, 1/2], [\,\,\,]],\frac{1}{4\ln(1/y)} \right) \\[0.7\baselineskip] \label{eqI33}
&=&  \frac{e/\sqrt{\pi}}{\ln(1/y)} \,\, G\left([[\,\,\,], [\,\,\,]], [[1/2], [\,\,\,]],\frac{1}{4\ln(1/y)} \right).
\end{eqnarray}
which, as expected, reproduces precisely Eq.~(\ref{eqI19}).

\noindent
2. $k=3$, $l=1$ and $\nu=0$:
\begin{eqnarray}\label{eqI34}
\widetilde{W}^{(0)}(3, 1; y) &=& \frac{e\sqrt{3}/(2\pi)}{\ln(1/y)} \,\, G\left([[\,\,\,], [0]], [[0, 1/3, 2/3], [\,\,\,]],\frac{1}{27\ln(1/y)} \right) \\[0.7\baselineskip]\nonumber
&=& \frac{e\sqrt{3}/(2\pi)}{\ln(1/y)} \,\, G\left([[\,\,\,], [\,\,\,]], [[1/3, 2/3], [\,\,\,]],\frac{1}{27\ln(1/y)} \right) \\[0.7\baselineskip]
&=& \frac{e}{3\pi}\,\left[\ln(1/y)\right]^{-3/2}\, K_{1/3}\left(\frac{2\sqrt{3}}{9[\ln(1/y)]^{1/2}}\right), \label{eqI35}
\end{eqnarray}
where $K_{\nu}(z)$ is the modified Bessel function of the second kind. Eq.~(\ref{eqI35}) can  also be checked with formula 3.13, p.~175 of \cite{FOberhettinger74}. With Eq.~(\ref{eqI31}) the corresponding spectrum varies as
\begin{equation}\label{eqI36}
e^{(1, 0)}(3, 1; n) \to 1 - \frac{1}{3\, n^{2/3}} + \ldots, \,\, n\to\infty,
\end{equation}
yielding $\sigma = 1/2$ and $c=1/3$.

\noindent
3. $k=3$, $l=1$ and $\nu=1/2$:
\begin{eqnarray}\nonumber
\widetilde{W}^{(1/2)}(3, 1; y) = \frac{e\sqrt{3}/(2\pi)}{[\ln(1/y)]^{1/2}} \,\, G\left([[\,\,\,], [1/2]], [[0, 1/3, 2/3], [\,\,\,]],\frac{1}{27\ln(1/y)} \right)
\end{eqnarray}
\begin{equation}\label{eqI38}
= \frac{e}{3(\pi)^{3/2}}\,\left[\ln(1/y)\right]^{-1}\, K_{1/3}\left(\frac{\sqrt{3}}{9[\ln(1/y)]^{1/2}}\right)\cdot K_{2/3}\left(\frac{\sqrt{3}}{9[\ln(1/y)]^{1/2}}\right),
\end{equation}
which yields the same $n$-dependence as in Eq.~(\ref{eqI36}).

\begin{figure}[!h]
\includegraphics[scale=0.4]{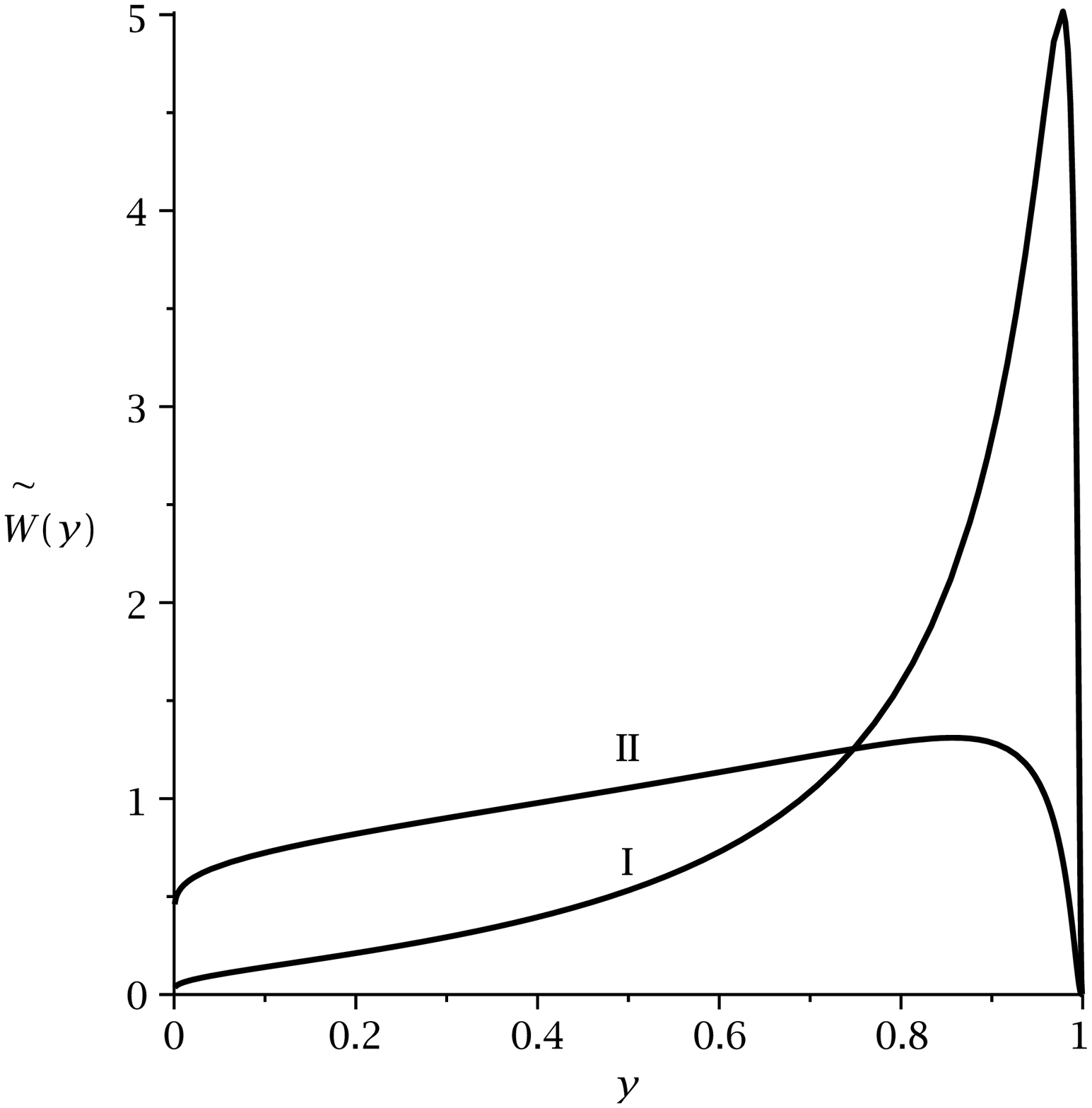}
\caption{\label{fig1} Plot of the weight functions $\widetilde{W}(y)$; the line $I$ corresponds to $\widetilde{W}^{(0)}(3, 1; y)$, see Eq.~(\ref{eqI35}) and the line $II$ represents $\widetilde{W}^{(1/2)}(3, 1; y)$, see Eq.~(\ref{eqI38}).}
\end{figure}

It is instructive to compare  weight functions leading to the same spectrum asymptotics. To this end we present the weight functions from Eqs.~(\ref{eqI36}) and~(\ref{eqI38}) in Fig.~\ref{fig1} .

\noindent
4. $k=3$, $l=2$ and $\nu=0$:\\
\begin{eqnarray}\nonumber
\widetilde{W}^{(0)}(3, 2; y) = \frac{e\, \sqrt{3/\pi}}{\ln(1/y)}\, G\left([[\,\,\,], [1/2]], [[1/3, 2/3], [\,\,\,]], \frac{4}{27[\ln(1/y)]^2}\right),
\end{eqnarray}
which can be neatly expressed in terms of modified Bessel functions $K_{\nu}(z)$:
\begin{eqnarray}\nonumber
\widetilde{W}^{(0)}(3, 2; y) &=& \frac{e\, 2\sqrt{3}}{27\pi}\,[\ln(1/y)]^{-3}\, \exp\left(\frac{2}{27 [\ln(1/y)]^2}\right)\, \\[0.7\baselineskip]\label{eqI39}
&\times& \left[K_{1/3}\left(\frac{2}{27[\ln(1/y)]^2}\right) + K_{2/3}\left(\frac{2}{27[\ln(1/y)]^2}\right) \right]
\end{eqnarray}
which leads to the asymptotics:
\begin{equation}\label{eqI39a}
e^{(1, 0)}(3,2; n) \to 1 - \frac{2}{3\, n^{1/3}} + \ldots,
\end{equation}
with $\sigma = 2/7$ and $c=2/3$.

\begin{figure}[!h]
\includegraphics[scale=0.4]{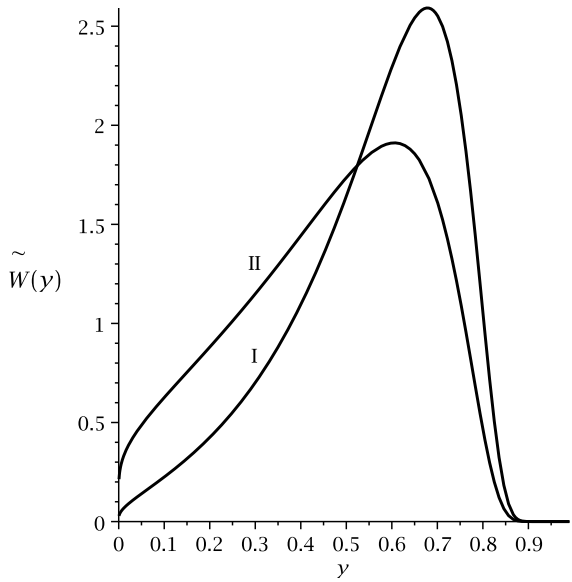}
\caption{\label{fig2} Plot of the weight functions $\widetilde{W}(y)$; the line $I$ corresponds to $\widetilde{W}^{(0)}(3, 2; y)$, see Eq.~(\ref{eqI39}) and the line $II$ represents $\widetilde{W}^{(1/4)}(3, 2; y)$, see Eq.~(\ref{eqI40}).}
\end{figure}

\noindent
5. $k=3$, $l=2$, $\nu = 1/4$:\\
This case can be expressed by the $G$ function
\begin{eqnarray}\nonumber
\widetilde{W}^{(1/4)}(3, 2; y) &=& \frac{2^{-1/4}\,\sqrt{3}}{[\ln(1/y)]^{3/4}}\, \\[0.7\baselineskip]\label{eqI40}
&\times& G\left([[\,\,\,], [1/8, 5/8]], [[0, 1/3, 2/3], [\,\,\,]], \frac{4}{27\,[\ln(1/y)]^2}\right),
\end{eqnarray}
which has a representation in terms of a sum of three hypergeometric functions of type $\,_{2}F_{2}$, which will not be quoted here. The asymptotics is that of Eq.~(\ref{eqI39a}). The weight functions from Eqs.~(\ref{eqI39}) and (\ref{eqI40}) which share the same spectrum asymptotics are compared in Fig.~\ref{fig2}.

\noindent
6. $k=4$, $l=1$, $\nu=0$:\\
The corresponding weight function $\widetilde{W}^{(0)}(4, 1; y)$ has an exact representation in terms of a sum of three hypergeometric functions of type $_{0}F_{2}$ which we will not given here.

This pattern extends to higher values of $k$ and $l$ for which the weight functions become increasingly complicated. They can however be fully handled analytically and graphically with a relative ease: we always deal with finite sums of hypergeometric functions.

We shall go over to further examples and employ a wealth of formulae of Ref. \cite{APPrudnikov92} and \cite{FOberhettinger74}, different from that of Eq.~(\ref{eqI24}).
\ \\

\noindent 
7. We now  choose  a function differing from the exponential which nevertheless yields  asymptotic behaviour  which is close to Eq. (\ref{eqI21}): this is provided by the formula 3.16.6.7, p. 358 of \cite{APPrudnikov92} or, alternatively by Eq. 7.69, p. 230 (for $a=1$) of \cite{FOberhettinger74}, namely:
\begin{eqnarray}\nonumber
\int_{0}^{1} &y^n& \left[\frac{1}{2\,K_{\nu}(1)}\, \frac{1}{\sqrt{\pi\ln(1/y)}} \, \exp(-\frac{1}{8\ln(1/y)})\, K_{\nu/2}\left(\frac{1}{8\ln(1/y)}\right)\right]\, dy = \\[0.7\baselineskip]\label{eqI41}
&=& \frac{1}{K_{\nu}(1)}\, \frac{K_{\nu}(\sqrt{n+1})}{\sqrt{n+1}} \equiv \rho_{\nu}^{(K)}(n), \,\, n=0, 1, \ldots\, .
\end{eqnarray}
The modified Bessel functions $K_{\nu}(z)$ in Eq.~(\ref{eqI41}) for $\nu\neq p/2$, $p=1, 2, \ldots$ are not elementary functions. The weight function in the l.h.s. of Eq.~(\ref{eqI41}) is positive on $[0, 1]$ and normalized as $\rho_{\nu}^{(K)}(0) = 1$. The asymptotic behaviour  for $\nu = 4/3$ is close to that of Eq.~(\ref{eqI21}):
\begin{eqnarray}\label{eqI42}
e^{(K)}(n) \sim 1 - \frac{1}{2 n^{1/2}} - \frac{1}{8 n} + \ldots,
\end{eqnarray}
leading to $c=1/2$ and $\sigma = 2/5$.

\begin{figure}[!h]
\includegraphics[scale=0.4]{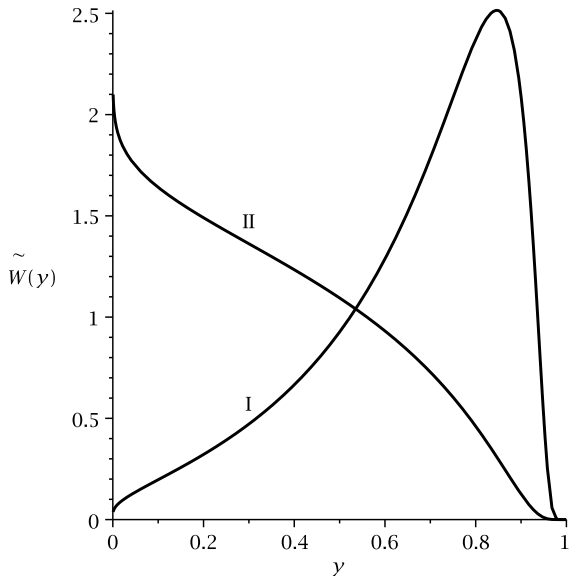}
\caption{\label{fig3} Plot of the weight functions $\widetilde{W}(y)$: the line $I$ corresponds to $\widetilde{W}^{(0)}(2, 1; y)$, see Eqs.~(\ref{eqI33}) and (\ref{eqI19}) and the line $II$ represents the weight function in Eq.~(\ref{eqI41}).}
\end{figure}

The weight functions from Eqs.~(\ref{eqI19}) or (\ref{eqI34}), and Eq.~(\ref{eqI43}) display the same spectrum asymptotics and are illustrated in Fig.~\ref{fig3}.

\noindent
8. We use now Eq. 2.2.2.1, p.~53 of \cite{APPrudnikov92} for the choice $\nu = 4/3$ and $a=1$, which leads to the normalized Hausdorff moment problem:
\begin{eqnarray}\label{eqI43}
\int_{0}^{1}\, y^n \, \left[e^{-1} \left[\ln(1/y)\right]^{1/6} I_{1/3}\left(2\sqrt{\ln(1/y)}\right) \right] \, dy \,=\, e^{-1}\, \frac{e^{1/(n+1)}}{(n+1)^{4/3}},
\end{eqnarray}
giving the asymptotics with $\sigma = 2/3$ and $c = 4/3$:
\begin{eqnarray}\label{eqI44}
e(n) \sim 1 - \frac{4}{3 n} + \ldots\, .
\end{eqnarray}
In Eq.~(\ref{eqI44}) $I_{\nu}(z)$ is the modified Bessel function of the first kind. Note that the value $\sigma = 2/3$ cannot be obtained from Eq.~(\ref{eqI32}).

\noindent
9. In this final example we address the problem of the one-dimensional Coulomb potential $\sim -|x|^{-1}$ for which the exact spectrum is \cite{LDLandau00}
\begin{eqnarray}\label{eqI45}
e_{c}(n) = 1 - \frac{1}{(n+1)^2}, \, n=0, 1, \ldots, \infty.
\end{eqnarray}
The  CS for this case have been constructed in \cite{JPGazeau99} and the corresponding moments are given by $\rho_{c}(n) =\frac{n+2}{2(n+1)}$, $n=0, 1, \ldots$, and they lead to the exact form for the corresponding weight function
\begin{eqnarray}\label{eqI46}
\widetilde{W}_{c}(y) = \frac{1}{2}\, \left[1 + \delta(y-1^{-})\right]\, H(1-y).
\end{eqnarray}
The Coulomb problem, even in its simplified version treated here, presents a particularity in that its resolution of unity is distributional in character as it involves, in Eq.~(\ref{eqI46}), the Dirac delta function $\delta(z)$. Although the solution Eq.~(\ref{eqI46}) is unique for the exact moments $\rho_{c}(n)$ defined above, we are able to construct another weight function $\widetilde{V}_{c}(y)$ which will asymptotically reproduce $e_{c}(n)$ of Eq. (\ref{eqI45}). To this end we use the formula 2.2.2.8, p. 54 of \cite{APPrudnikov92} for $a=1$, which gives
\begin{eqnarray}\nonumber
\int_{0}^{1} &y^n& \widetilde{V}_{c}(y)\, dy \,=\, \int_{0}^{1}\, y^n\, \left[e^{-1} \left(\frac{I_{1}(2\sqrt{\ln(1/y)})}{[\ln(1/y)]^{1/2}} + \delta(y - 1^{-})\right) \right]\, dy \\[0.7\baselineskip]\label{eqI47}
&=& e^{-1}\, e^{1/(n+1)}, \,\,n=0, 1, \ldots,
\end{eqnarray}
with asymptotics $\tilde{e}_{c}(n) \to 1 - n^{-2} + 3 n^{-3} + \ldots\,\,$. Note that $\widetilde{V}_{c}(y)$ displays a non-trivial dependence on $y\in[0, 1]$ but still retains a Dirac peak at $y=1$. We are not aware of existence of any weight function without Dirac's delta leading to Eq.~(\ref{eqI45}).

\begin{figure}[!h]
\includegraphics[scale=0.4]{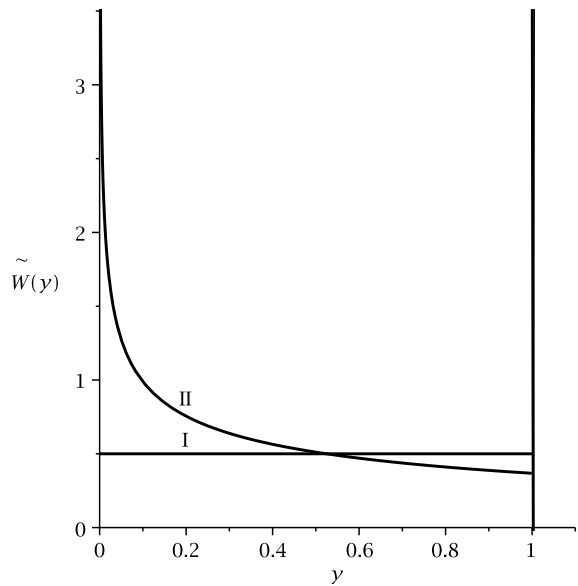}
\caption{\label{fig4} Plot of the weight functions $\widetilde{W}(y)$: the line $I$ corresponds to $\widetilde{W}_{c}(y)$, see Eqs.~(\ref{eqI46}) and the line $II$ represents the weight function $\widetilde{V}_{c}(y)$, see Eq.~(\ref{eqI47}). The delta peak at $y=1$ are represented by a vertical line. Observe that this is a schematic plot: different coefficients ($1/2$ and $e^{-1}$, s. Eq.~(\ref{eqI46}) and~(\ref{eqI47}) respectively) in front of delta peaks at $y=1$ cannot be accounted for.}
\end{figure}

The different weight functions relevant for the Coulomb interaction, both of them involving Dirac's delta function at $y=1$, are schematically displayed on Fig.~\ref{fig4}.


\section{Discussion and Conclusion}

In this work we  used a semiclassical approximation for the spectra of one-dimensional inverse power-law potentials to construct approximate coherent states, relevant for these potentials.  Our goal in using this construction was to achieve the resolution of unity for a given potential characterized by an exponent $\sigma$ in $V_{\sigma}(x)\sim -|x|^{-\sigma}$. In this sense the $\rho_{\sigma}(n)$'s of Eq.~(\ref{eqI9}) should be perceived as a sort of trial parameters which should asymptoticaly reproduce $e_{\sigma}(n)$ via Eq.~(\ref{eqI10}). This leads, of course, to a multiplicity of possible choices of $\rho_{\sigma}(n)$ leading to the same $e_{\sigma}(n)$. The price of this approximation is the fact, that the temporal stability characterizing exact Gazeau-Klauder CS \cite{JPGazeau99, JBenGeloun09} denoted by $|J, \gamma, \sigma\rangle_{G K}$ (i.e. the equality $e^{-i h_{\sigma} t} |J, \gamma, \sigma\rangle_{G K} \,=\, |J, \gamma + t, \sigma\rangle_{G K}$) will be only approximately satisfied by the states of Eq.~(\ref{eqI6}). On the contrary, the basic Gazeau-Klauder axiom of the resolution of unity is fully maintained in our construction.  

For certain values of $\sigma$, more precisely for rational $\sigma$ such that $0<\sigma\leq 2/3$ and for $\sigma=1$, we are able to produce many resolutions of unity that are asymptotically relevant for one and the same $\sigma$. This is due in the first place to two key formulae, Eq.~(\ref{eqI24}) and~(\ref{eqI32}) which involve the use of the inverse Laplace transform. For other values of $\sigma$ in the range $2/3<\sigma<2$ (except $\sigma = 1$) our approach does not produce any required solutions for the resolution of unity.

We should comment here on the nature of approximation involved in the formulation of Eq.~(\ref{eqI6}). Since for general $\sigma$ and $n$ neither exact spectra nor exact eigenfunctions $|n, \sigma\rangle$ are known we strike a compromise in Eq.~(\ref{eqI6}) by retaining the exact orthonormal eigenfunctions and replacing the energy spectrum by its quasi-classical form of Eq.~(\ref{eqI4}). 

In Eq.~(\ref{eqI6}) we substitute the quasiclassical approximation for $e_{\sigma}(n)$ instead of the exact spectrum. This means that we are using the asymptotic approximation in the low energy region, where its use is not \textit{a priori} justified. However, for certain values of $\sigma$ the quasiclassical approximation is very successful even down to the low energy: in fact, for the linear repulsive potential $V(x)\sim x$ ($x\geq 0$; $V(x) = \infty$, $x< 0$) it predicts the correct energies and wave functions right down to the ground state \cite{CQuigg79, VGalitski85}. For attractive potentials, for which in general one does not have exact solutions, the agreement is less spectacular \cite{VGalitski85} but in general the quasiclassical approximation works well in large parts of the spectra and not only for $n\to\infty$, in which region it tends to the exact solution. Therefore we believe that the use of Eq.~(\ref{eqI4}) in Eq.~(\ref{eqI6}) is a reasonable prescription.


\section{Acknowledgements}

The authors acknowledge support from Agence Nationale de la Recherche (Paris, France) under Program No. ANR-80-BLAN-0243-2 and from PAN/CNRS Project PICS No.4339 (2008-2010).


\appendix
\section{}

We give here a streamlined proof of positivity $\widetilde{W}^{(1, \nu)}(k, l; y)$, for $0\leq y\leq 1$ from Eq.~(\ref{eqI29}) under the conditions $k$, $l$ positive integers, $k>l$, and $\nu\geq 0$.

Consider first Meijer's G function of Eq.~(\ref{eqI24}), with $\Delta(k, a)~=~\frac{a}{k}, \, \frac{a+1}{k},\, \ldots, \, \frac{a+k-1}{k}$:
\begin{equation}\label{eqA1}
G\left([[\,\,\,], [\Delta(l, \nu)]], [[\Delta(k, 0)], [\,\,\,]], z\right) = \mathcal{M}^{-1}\left[ \frac{\prod_{j=0}^{k-1}\Gamma(s + \frac{j}{k})}{\prod_{j=0}^{l-1}\Gamma(s + \frac{\nu+j}{l})}; z\right]
\end{equation}
\begin{equation}\label{eqA2}
\qquad\qquad\qquad\qquad = \mathcal{M}^{-1}\left[\left(\prod_{j=0}^{l-1}\frac{ \Gamma(s +
\frac{j}{k})}{ \Gamma(s+\frac{\nu+j}{l})}\right) \,\prod_{j=l}^{k-1} \Gamma(s + \frac{j}{k}); z \right].
\end{equation}
The convolution property for $\mathcal{M}[f(x); s] = f^{*}(s)$ and $\mathcal{M}[g(x); s] = g^{*}(s)$
\begin{equation}\label{eqA3}
\mathcal{M}^{-1}\left[f^{*}(s)\, g^{*}(s); x \right] \,=\, \int_{0}^{\infty} f(x/t)\, g(t)\, \frac{d t}{t} \,=\, \int_{0}^{\infty} g(x/t)\, f(t)\, \frac{d t}{t}
\end{equation}
for $f(x)>0$ and $g(x)>0$ clearly conserves the positivity, see \cite{KAPenson10} and references therein. Fix $\nu\geq 0$ and consider $\mathcal{M}^{-1}$ of an individual term of the first product in (\ref{eqA2}). It will obey, using the formula 8.4.2.3, p. 631 of \cite{APPrudniko98v}
\begin{equation}\label{eqA4}
\mathcal{M}^{-1}\left[\frac{\Gamma(s + \frac{j}{k})}{\Gamma(s + \frac{\nu+j}{l})}; x\right] \,=\, \frac{x^{j/k}\,(1-x)^{-1 + \frac{1}{l}\left(\nu + j\,\frac{k-l}{k}\right)}}{\Gamma\left(\frac{1}{l}\left(\nu + j\,\frac{k-l}{k}\right)\right)}, \,\,\, j=0, 1, \ldots, l-1,
\end{equation}
which is a positive function for $0<x<1$. Also, concerning the second product in (\ref{eqA2}) an individual term evidently gives
\begin{equation}\label{eqA5}
\mathcal{M}^{-1}\left[\Gamma\left(s + \frac{j}{k}\right); x\right] \,=\, x^{j/k}\, e^{-x} \,>\,0, \,\,\, x>0, \,\,\, j = l, \ldots, k-1.
\end{equation}
Then (\ref{eqA2}) can be viewed as a multiple, $k$-fold Mellin convolution of positive functions, which by (\ref{eqA3}) is itself positive.

The proof of positivity of $\widetilde{W}^{(1, \nu)}(y)$ is completed by remarking that if $F(y)$ is positive on $[0, \infty)$ then for $\alpha$ and $\beta$ arbitrary $\left[\ln\left(\frac{1}{y}\right)\right]^{\alpha}\, F\left(\left[\ln\left(\frac{1}{y}\right)\right]^{-\beta}\right)$ is positive on $[0, 1]$. We stress that the $\nu<0$ case destroys the positivity of $\widetilde{W}^{(1, \nu)}(y)$ and is not relevant for our purposes.

\end{document}